\documentclass{article}

\def\beq{\begin{eqnarray}}
\def\eeq{\end{eqnarray}}

\begin{document}

\title{Divergencies in the Casimir energy for a
medium with realistic ultraviolet behavior }
\author{H. Falomir \\
IFLP - Dept. de F{\`\i}sica, Fac. de Ciencias Exactas,\\
Universidad Nacional de La Plata, \\ C.C. 67, (1900) La Plata, Argentina\\
$\phantom{a}$\\
K. Kirsten\\
Department of Physics and Astronomy, \\ The University of
Manchester, Theory Group,\\  Schuster Laboratory,  Manchester M13
9PL, England \\$\phantom{a}$\\
and K. R{\'e}bora\\
IFLP - Dept. de F{\`\i}sica, Fac. de Ciencias Exactas,\\
Universidad Nacional de La Plata, \\ C.C. 67, (1900) La Plata,
Argentina}


\maketitle

\begin{abstract}
We consider a dielectric medium with an ultraviolet behavior as
it follows from the Drude model. Compared with dilute models,
this has the advantage that, for large frequencies, two different
media behave the same way. As a result one expects the Casimir
energy to contain less divergencies than for the dilute media
approximation. We show that the Casimir energy of a spherical
dielectric ball contains just one divergent term, a volume one,
which can be renormalized by introducing a contact term analogous
to the volume energy counterterm needed in bag models.


\end{abstract}

\vspace{1,5cm}

Due to recent progress in experimental techniques, it is now
possible to measure Casimir forces between macroscopic testbodies
very accurately \cite{lamo97-78-5,mohi98-81-4549}. Even quantum
mechanical actuation of microelectromechanical systems by the
Casimir force is under discussion \cite{chan01-8feb-1}. For these
reasons a thorough theoretical understanding of the Casimir effect
with realistic media is desirable. There are essentially two ways
to calculate the Casimir energy. One possibility is to sum up
retarded van der Waals forces between individual molecules
\cite{casi48-73-360,land60b}. The second way makes use of quantum
field theory in the background of a dielectric medium (see
e.g.~\cite{schw78-115-1,milt80-127-49,cand82-143-241,baak86-30-413,plun86-134-87,most97b,milt99pre,brev94-27-6853}).
In this second approach one tries to recover retarded van der
Waals forces by calculating the vacuum energy for the
electro-magnetic field in a dielectric background. The relation
between these two approaches is not well established. Only for a
dilute ball, up to the second order of a perturbative expansion,
the methods have been shown to yield the same answers
\cite{brev99-82-3948,bart99-32-525}. In all other cases of
dielectric media the presence of divergencies forbids to give a
physical interpretation to the finite parts obtained. Even for a
dilute medium, once the full dependence of the pole on the
dielectric constants is considered, the result is not even
expressible in terms of known special functions
\cite{bord99-59-085011}. So the counterterms needed to
renormalize the divergent contributions is extremely complicated
and the interpretation of the classical model associated with the
couterterms is completely unclear.

One possible reason for this briefly described difficulty might be
the fact that the models of dielectric media treated in the
quantum field theory framework mostly do not fulfill a realistic
frequency dependent dispersion relation. A direct consequence of
an unreasonable ultraviolet behavior is, for example, the
appearance of divergencies which make it impossible to extract, in
a physically reasonable way, a finite value for the Casimir
energy.

For those reasons, it seems natural to analyze the divergencies of
the Casimir effect in a model with a dispersion relation that
shows realistic features, at least in the ultraviolet range. As
we will see, once this behavior is assumed, the pole structure
resulting in the $\zeta$-function regularization scheme is in fact
very simple. Namely just one pole exists and it is a term
completely analogous to the one renormalized via the bag constant
in the bag model. So, introducing only one counterterm, namely a
volume energy, the Casimir energy can be rendered finite.
Apparently there is no intrinsic way within the field theory
framework, to fix the finite part of this counterterm. In
principle, as is the case in bag models, it has to be fixed by
experiments. However, given the simple structure of the
renormalization needed, we feel this is an important step in
order to understand the relation between different approaches
employed and to extract finite Casimir energies in realistic
media.

So, let us start with a description of the simple model we are
going to analyze. We will consider a nonmagnetic ($\mu=\mu_0$)
dielectric  ball of radius $a$, with a frequency dependent
permittivity given by $\epsilon (\omega) = \epsilon_{0}
(1-\frac{\Omega^{2}}{\omega^{2}})$, immersed in another medium of
permittivity $\epsilon_{0}$. This frequency dependent
permittivity is the so-called plasma model, which follows from the
Drude model in the high frequency approximation. The
phenomenological parameter $\Omega$ is usually referred to as the
effective plasma frequency. Given that we are going to
concentrate only on the pole structure of the Casimir energy for
the described setting, this high frequency model is sufficient for
our purposes. Formally, the Casimir energy of this configuration
is defined by summing over the vacuum energies of each mode of the
electromagnetic field,
\begin{equation}\label{sum-cas}
  E_{Cas}(a) =  \sum_{n} \
   \frac{1}{2} \,
  \hbar \, \omega_{n} =\frac{\hbar\, c}{2 a}  \,
  \sum_{n} \, z_{n},
 \end{equation}
where the dimensionless quantities $z_{n} = {a \omega_{n}}/{c}$
are the eigenvalues associated with a radius one ball. As it
stands, the sum in Eq.\ (\ref{sum-cas}) is divergent, and a
regularization procedure must be adopted. To properly define the
Casimir energy we will use the zeta function regularization. In
this scheme, $E_{Cas}(a)$ is defined as the analytic continuation
of the function \beq E_{Cas}(a) \equiv \left. \frac{\hbar c}{2a}
\zeta (s)\right|_{s\rightarrow -1}, \label{defcasi} \eeq where
$\zeta(s)$ is defined through the eigenfrequencies of the
electromagnetic field by means of the series \beq
\label{zeta-def} \zeta (s) = \sum_n z_n^{-s} , \eeq which is
absolutely and uniformly convergent for $\Re (s)$ sufficiently
large. As is well known, usually the zeta function has a pole at
$s=-1$, in which case a suitable renormalization procedure has to
be employed.

In Eq.\ (\ref{zeta-def}), the eigenfrequencies $\omega_{n}$ must
be determined by solving the field equation
\begin{equation}\label{Helmholtz}
\triangle \overrightarrow{E} + \mu \epsilon
\frac{\omega^2}{c^2}\overrightarrow{E}=0
\end{equation}
(and similarly for $\overrightarrow{B}$), subject to matching
conditions appropriate to the interphase between two dielectric
media,
\begin{equation}\label{BC}
 \left.\phantom{\frac{1}{\mu_2}} E_{\theta, \phi}\right|_{r=a^+} =
 \left. \phantom{\frac{1}{\mu_2}} E_{\theta, \phi}\right|_{r=a^-},\
\left. \frac{1}{\mu_1}B_{\theta, \phi}\right|_{r=a^+} =
\left.\frac{1}{\mu_2}B_{\theta, \phi}\right|_{r=a^-}.
\end{equation}

One can consider the transversal electric (TE) modes, for which
the electric field has the form
\begin{equation}\label{TE}
  \overrightarrow{E}_{l,m}=f_l(r) \overrightarrow{L}
  Y_{l,m}(\theta,\phi),
\end{equation}
and separately the transversal magnetic modes (TM), with the
magnetic field given by
\begin{equation}\label{TM}
  \overrightarrow{B}_{l,m}=g_l(r) \overrightarrow{L}
  Y_{l,m}(\theta,\phi).
\end{equation}
Here, as usual,
\begin{equation}\label{L}
  \overrightarrow{L}=-\imath \overrightarrow{r}\times
  \overrightarrow{\nabla} .
\end{equation}
The imposition of the conditions in Eq.\ (\ref{BC}) leads to
\begin{equation}\label{BC-TE}\begin{array}{c}
  f_l(r=a^+)=f_l(r=a^-), \\ \\
  \left. \ \partial_r
\left[r f_l(r)\right]\right|_{r=a^+}= \left. \
\partial_r \left[r f_l(r)\right]\right|_{r=a^-},
\end{array}
\end{equation}
for the TE modes, and
\begin{equation}\label{BC-TM}\begin{array}{c}
 g_l(r=a^+)=  g_l(r=a^-), \\ \\\displaystyle{
  \left. \frac{1}{ \epsilon_0}\ \partial_r
\left[r g_l(r)\right]\right|_{r=a^+}}=\displaystyle{ \left.
\frac{1}{\epsilon (\omega)}\
\partial_r \left[r g_l(r)\right]\right|_{r=a^-}},
\end{array}
\end{equation}
for the TM modes.

In order to have a discrete spectrum, one might enclose the
system into a large conducting sphere of radius $R$, sending
$R\to\infty$ at a suitable intermediate step. This implies also
the following boundary condition at $r = R$ for $f_l(r)$ and
$g_l(r)$:
\begin{equation}\label{en R}
  \left. f_l(r)\right|_{r=R}=0,\qquad  \left.\partial_r
  \left(r\, g_l(r)\right)\right|_{r=R}=0.
\end{equation}
Alternatively it is possible to use a formulation in terms of the Jost function
of the corresponding scattering problem, where a subtraction of the Minkowski
space contribution is performed at the beginning and the infinite volume
limit taken implicitly \cite{bord99-59-085011}.

It is straightforward to see that the eigenfrequencies for TE
modes are determined by the zeroes of the function (we put $\nu = l+1/2$)
\begin{equation}\label{deltaTE-1}\begin{array}{c}
  \Delta^{TE}_{\nu} (z)
  =\mathcal{J}_{\nu}(\bar{z}_1)\left\{
\mathcal{Y}_{\nu}(\bar{z}_0) \mathcal{J}'_{\nu}(\bar{z}_2)
    -\mathcal{J}_{\nu}(\bar{z}_0)
    \mathcal{Y}'_{\nu}(\bar{z}_2) \right\} - \\ \\
  - \xi \, \mathcal{J}'_{\nu}(\bar{z}_1) \left\{
\mathcal{Y}_{\nu}(\bar{z}_0) \mathcal{J}_{\nu}(\bar{z}_2)
    -\mathcal{J}_{\nu}(\bar{z}_0)
    \mathcal{Y}_{\nu}(\bar{z}_2) \right\},
\end{array}
\end{equation}
where
\begin{equation}\label{funciones-1}\begin{array}{c}
  \mathcal{J}_{\nu}(w)=w\ j_l(w) =\displaystyle{
  \sqrt{\frac{\pi w}{2}}} J_{\nu}(w) \\ \\
  \mathcal{Y}_{\nu}(w)=w\ y_l(w) =\displaystyle{
  \sqrt{\frac{\pi w}{2}}} Y_{\nu}(w)
\end{array}
\end{equation}
are the Riccati - Bessel functions, being $z=a\,(\omega/c)$,
$\bar{z}_{1}=z \, n(z)$, $\bar{z}_{2}=z \, n_{0}$,
$\bar{z}_{0}=z\, R \, n_{0}/a $, $\xi= n(z)/n_{0}$, $n_{0} =
\sqrt{\epsilon_{0}}$, and $n(z) = \sqrt{\epsilon (\omega)}$.

Similarly, for the TM modes the eigenfrequencies are determined
by the zeroes of
\begin{equation}\label{deltaTM-1}\begin{array}{c}
  \Delta^{TM}_{\nu} (z)
  =\mathcal{J}_{\nu}(\bar{z}_1)\left\{
\mathcal{Y}'_{\nu}(\bar{z}_2) \mathcal{J}'_{\nu}(\bar{z}_0)
    -\mathcal{J}'_{\nu}(\bar{z}_2)
    \mathcal{Y}'_{\nu}(\bar{z}_0) \right\} - \\ \\
  - \displaystyle{\frac{1}{\xi}}\, \mathcal{J}'_{\nu}(\bar{z}_1) \left\{
\mathcal{Y}_{\nu}(\bar{z}_2) \mathcal{J}'_{\nu}(\bar{z}_0)
    -\mathcal{J}_{\nu}(\bar{z}_2)
    \mathcal{Y}'_{\nu}(\bar{z}_0) \right\}.
\end{array}
\end{equation}
We will suppose that these functions have only real and simple
zeroes in the open right half plane.

\bigskip

In the following we will consider explicitly only the TE modes,
since the treatment of the TM modes is entirely similar. Due to
the spherical symmetry of the problem, the $\zeta$-function has
the general appearance \beq \zeta (s) = \sum_{\nu=3/2}^\infty 2
\nu \sum_n z_{\nu , n}^{-s} , \label{defzeta} \eeq where $\nu$
labels the angular momentum and, for a given $\nu$, the index $n$
labels the zeroes of $ \Delta^{TE}_{\nu} (z)$ in
Eq.~(\ref{deltaTE-1}). We need to construct the analytic
continuation of this function to $s\approx -1$. As we will see,
$\zeta (s)$ has a simple pole at $s=-1$ and $E_{TE}(a)$, defined as
in Eq.\ (\ref{defcasi}), contains a divergent term which  depends
on the particular dispersion relation adopted for $\epsilon
(\omega)$. The method employed for the analytic continuation has
been explained in great detail, e.g., in
\cite{bord96-37-895,bord99-59-085011,eliz93-26-2409,lese94-27-2483,falo01-63-025015},
and we will simply follow this procedure. However, let us mention
that the idea of applying this method in the specific context of
dielectric media goes back at least to \cite{kamp68-26-307}.
\bigskip

For $\Re (s)$ large enough, we can represent the $\zeta$-function
as an integral in the complex $z$-plane  employing the Cauchy's
theorem. For the TE modes we have
\begin{equation}\label{integral}
  \zeta_\nu(s) := \sum_{n=1}^{\infty} \ z_{\nu,n}^{-s} =
  \displaystyle{\frac{1}{2\pi \imath}} \oint_C z^{-s}
  \displaystyle{\frac{{\Delta^{TE}_\nu}' (z)}{\Delta^{TE}_\nu (z)}} \  dz,
\end{equation}
where the curve $C$ encloses counterclockwise all the positive
zeros of $\Delta^{TE}_\nu (z)$. This curve can be deformed into a
straight vertical line, crossing the horizontal axis at $\Re (z) =
x$, where $x$ is any value satisfying $0<x<z_{\nu,1}$,
$z_{\nu,1}$  being the first positive zero of $\Delta^{TE}_\nu
(z)$.  Obviously the integral in eq. (\ref{integral}) does not
depend on the particular value of $x$ in this range.

Indeed, expressing the integrand in terms of the modified Bessel
functions  and taking into account their asymptotic behavior for
large arguments, it is easily seen that the integral
\begin{equation}\label{zeta}
    \zeta_\nu(s) =
  \displaystyle{\frac{-1}{2\pi \imath}}
  \int_{-\imath\infty+x}^{\imath\infty+x } z^{-s}
  \displaystyle{\frac{{\Delta^{TE}_\nu}' (z)}{\Delta^{TE}_\nu (z)}} \  dz,
\end{equation}
converges absolutely and uniformly to an analytic function in the
open half-line $s > 1$, which can be meromorphically extended to
the whole complex $s$-plane.

Expression (\ref{zeta}) can also be written as
\begin{equation}\label{zeta-final}\begin{array}{c}
  \zeta_\nu(s)= \displaystyle{-\frac{1}{\pi}}\,\Re \left\{ x^{1-s}\,
  e^{\displaystyle{-\imath \frac{\pi}{2}s}}
   \displaystyle{\int_0^\infty} (y-\imath)^{-s}\,
  \displaystyle{\frac{{\Delta^{TE}_\nu}'
  \left(\imath x(y-\imath)\right)}
  {\Delta^{TE}_\nu \left(\imath x(y-\imath)\right)}} \ dy
  \right\},
\end{array}
\end{equation}
where the prime means derivative with respect to the argument,
and we have made use of the properties of the Bessel functions of
complex argument.

Changing the integration variable to $t \equiv z (y - \imath)$,
with $z = x/\nu >0$, we finally get
\begin{equation}\label{zeta-final-2}\begin{array}{c}
  \zeta_\nu(s)= \displaystyle{-\frac{1}{\pi}}\,\Re \left\{ \nu^{-s}\,
  e^{\displaystyle{-\imath \frac{\pi}{2}(s+1)}}
   \displaystyle{\int_{- \imath z}^{\infty - \imath z}} t^{-s}\,
  \displaystyle{\frac{d( \ln \Delta^{TE}_\nu
  \left(\imath \nu t\right))}
  {d t}} \ dt
  \right\}.
\end{array}
\end{equation}

Notice that the right hand side of Eq.~(\ref{zeta-final-2}) does
not depend on $z$ for $z(>0)$ small enough.
\smallskip

In order to construct the analytic extension of the expression in
Eq.\  (\ref{defcasi}) to $s \approx -1$, we subtract and add the
first few terms of the asymptotic expansion of the integrand in
(\ref{zeta-final-2}) (obtained from the uniform asymptotic Debye
expansion for the modified Bessel functions appearing in
$\Delta^{TE}_\nu \left(\imath \nu t\right)$). In fact, in order
to isolate the singularities of the Casimir energy it is
sufficient to retain in this expansion terms up to the order
$\nu^{-3}$:
\begin{equation}\label{DD}
\frac{d( \ln \Delta^{TE}_\nu
  \left(\imath \nu t\right))}{d t}
  = D_{\nu}^{TE}(t)
   + \mathcal{O}(\nu^{-4}),
\end{equation}
with
\begin{equation}\label{Debye-expTE}\begin{array}{c}
   \displaystyle{ D_{\nu}^{TE}(t)= \nu D_{TE}^{(1)}(t) +
   D_{TE}^{(0)}(t) + \nu^{-1} D_{TE}^{(-1)}(t) + \nu^{-2} D_{TE}^{(-2)}(t)
   + \nu^{-3} D_{TE}^{(-3)}(t)},
\end{array}
\end{equation}
where $D_{TE}^{(k)}(t)$, $k = 1,...,-3$, are algebraic functions
of $t$ given in the Appendix. Notice that we have also discarded
terms which are exponentially vanishing for $R\rightarrow \infty$.

\bigskip

So, we must consider the series
\begin{equation}\label{sustrac}\begin{array}{c}
   \displaystyle{\sum_{\nu=3/2}^{\infty} \nu} \
   \Re \left\{\left. \displaystyle{
   \frac{-\nu^{-s}}{\pi}}\,
  e^{\displaystyle{-\imath \frac{\pi}{2}(s+1)}}
   \displaystyle{\int_{- \imath z}^{\infty - \imath z}} t^{-s}\,
  \displaystyle{\frac{d( \ln \Delta^{TE}_\nu
  \left(\imath \nu t\right))}
  {d t}} \ dt \
  \right\} \right. = \\ \\
    -\displaystyle{\sum_{\nu=3/2}^{\infty} \nu \,} \
    \Re \left\{ \left. \displaystyle{
   \frac{\nu^{-s}}{\pi}}\,
  e^{\displaystyle{-\imath \frac{\pi}{2}(s+1)}}
   \displaystyle{\int_{- \imath z}^{\infty - \imath z}} t^{-s}\,
  \displaystyle{D_{\nu}^{TE}(t)} \ dt \
  \right\} \right. - \\ \\
   \displaystyle{\sum_{\nu=3/2}^{\infty} \nu} \
   \Re \left\{ \left. \displaystyle{
   \frac{\nu^{-s}}{\pi}}\,
  e^{\displaystyle{-\imath \frac{\pi}{2}(s+1)}}
   \displaystyle{\int_{- \imath z}^{\infty - \imath z}} t^{-s}\,
 \left\{ \displaystyle{\frac{d( \ln \Delta^{TE}_\nu
  \left(\imath \nu t\right))}
  {d t} - D_{\nu}^{TE}(t)} \right\} \ dt \
  \right\}  \right. .
\end{array}
\end{equation}
The  second term in the right hand side of (\ref{sustrac})
converges for $s>-2$ by construction. Therefore, we can put $s=
-1$ inside the sum and the integral, and numerically evaluate
this contribution when necessary.

In order to investigate the ultraviolet divergencies in $E_{Cas}$
it is sufficient to consider the first term in the right hand
side of (\ref{sustrac})\footnote{\label{foot} It should be
stressed that the real part in the argument of the series in the
first term in the right hand side of Eq.\ (\ref{sustrac}) is in
fact independent of $z$: Taking into account the analyticity of
the integrand, one can investigate the $z$-dependence by studying
the integral
\begin{equation}\label{integ-exac}
    \displaystyle{\int_{- \imath z}^{1}} t^{-s}\,
  \displaystyle{D_{\nu}^{TE}(t)} \ dt,
\end{equation}
which can be exactly solved in terms of hypergeometric functions.
It is straightforward to verify that the $z-$dependent part is
imaginary for all $s>1$, and therefore is dropped out when taking
the real part in Eq.\ (\ref{sustrac}). This feature will be
useful in what follows.}.

\bigskip

For $s > 1$ we can study each term in $D_{\nu}^{TE}(t)$
separately, and evaluate the expressions
\begin{equation}\label{ordenes}\begin{array}{c}
   \displaystyle{\sum_{\nu=3/2}^{\infty} \nu^{-(s-k-1)}} \
   \Re \left\{ \displaystyle{
    \frac{- e^{\displaystyle{-\imath \frac{\pi}{2}(s+1)}}}{\pi}}\,
      \displaystyle{\int_{- \imath z}^{\infty - \imath z}} t^{-s}\,
  \displaystyle{D_{TE}^{(k)}(t)} \ dt
  \right\},
\end{array}
\end{equation}
with the index $k = 1,...,-3$ corresponding to the order in the
Debye expansion.

Since the real part of the expression between brackets in
(\ref{ordenes}) is independent of $\nu$ (it is independent of
$z=x/\nu$ - see footnote \ref{foot}), the sum over $\nu$ can be
performed by means of the Hurwitz zeta function, \beq
\sum_{\nu=3/2}^{\infty} \nu^{-(s-k-1)} = \zeta_H (s-k-1,1/2)
-2^{s-k-1}. \eeq Only for $k = -3$ one gets a singularity at $s =
-1$ given by
\begin{equation}\label{zeta-k=3}
\zeta_{H}(s+2,1/2)_{|_{s\simeq-1}} = \frac{1}{s + 1} + (\gamma
+\ln(4)) + \mathcal{O}(s+1) .
\end{equation}
For the other values of $k$ a regular extension to $s = -1$ is
obtained.

Therefore, it is only for $k=-3$ that we need to calculate both
finite and singular parts of the analytic extension of the
integral in Eq.\ (\ref{ordenes}) around $s = -1$, while for the
other values of $k$ only the singular terms are needed.

\bigskip

All these terms can be worked out exactly to give for the
divergent part of the contribution of the TE modes to the Casimir
energy (in the limit $R\to \infty$),
\begin{equation}\label{resultado-TE}
 E_{TE} (a) = \frac{\hbar c}{a} \,\left\{-\frac{
{n_{0}}^3\,a^4 \Omega^4  }
  {24\,\pi \,c^4\,\left( 1 + s \right) } -
\frac{n_0 a^2\Omega^2}{4\pi c^2(1+s)} + \mathcal{O} (s+1)^0
\right\}.
\end{equation}
Notice that the first term in the right hand side corresponds to
a volume contribution, while the second one is a curvature
contribution. In agreement with \cite{MP-V}, no surface
contribution appears.

For the TM modes one gets similarly
\begin{equation}\label{resultado-TM}
E_{TM}(a) = \frac{\hbar c}{a}  \, \left\{-\frac{
\,{n_{0}}^3\,a^4\,{\Omega }^4  }
  {24\,\pi\,c^4\,\left( 1 + s \right)} +
\frac{n_0 a^2\Omega^2}{4\pi c^2(1+s)} + \mathcal{O} (s+1)^0
\right\}.
\end{equation}

Finally, adding up the contributions coming from the TE and TM
modes, we get for the singular piece of the Casimir energy of the
dielectric ball just a volume contribution, \beq E (a) =
\frac{\hbar c}{a} \, \left\{-\frac{ {n_{0}}^3\,a^4 \Omega^4  }
{12\pi c^4 (1+s)}
 + \mathcal{O} (s+1)^0 \right\},
\eeq since the divergent curvature terms have canceled out between
TE and TM modes. Replacing the volume of the ball,  $V=4\pi
a^3/3$, we find \beq E(a) = -\frac{-\hbar n_0^3 V
\Omega^4}{16\pi^2 c^3 (1+s)}  + \mathcal{O} (s+1)^0. \eeq

Thus, we have found that the pole structure of the Casimir energy
in the framework of the $\zeta$-function regularization scheme is
very simple, once a realistic behavior of the dielectric medium at
high frequencies is assumed. The pure volume divergence might be
seen to represent a contribution to the mass density of the
material \cite{brev99-82-3948} and it is to be absorbed into a
phenomenological counterterm, much in the way it is done in bag
models. Neither a surface tension nor a curvature counterterm is
needed in this model, since divergencies nicely cancel out
between TE and TM modes.

Given this simple pole structure it makes sense to analyze further
finite parts of the Casimir energy in a quantum field theory
context for realistic media. As a first step one might assume the
Drude model for the medium and analyze how the Casimir energy
depends on e.g.~the plasma and the relaxation frequency. More
ambitiously, the dielectric constants might be obtained from
tabulated refractive indices using the Kramer-Kronig relation
\cite{klim00-61-062107}. This is under consideration and will
allow a detailed analysis of the Casimir energy for a spherically
symmetric situation with realistic media.
\\[.3cm]

{\bf Acknowledgements:} We thank C.\ G.\ Beneventano, S.\ Dowker,
M.\ de Francia and E.\ M.\ Santangelo for interesting
discussions.

Part of this work has been performed during a very pleasant stay
of K.\ K.\  at the Department of Physics of the University of La
Plata. He thanks all the members of this Department for their
very kind hospitality.

K.\ K.\ has been supported by the EPSRC under grant number
GR/M45726. H.\ F.\ and K.\ R.\ acknowledge the support received
from ANPCyT (PICT'97 Nr.\ 00039), CONICET (PIP Nr.\ 0459), and
UNLP (Proy.\ 11-X230), Argentina.

\appendix \label{Ds}

\section{Debye expansion for the TE and TM modes}

The functions $D_{TE}^{(k)}(t)$, $k = 1,...,-3$, coefficients of
the expansion $ D_{\nu}^{TE}(t)$ (see Eqs. (\ref{DD}) and
(\ref{Debye-expTE})), which are obtained from the uniform
asymptotic Debye expansion for the modified Bessel functions
appearing in $\Delta^{TE}_\nu \left(\imath \nu t\right)$, are
given by
\begin{equation}\label{DTE1}\begin{array}{c}
D_{TE}^{(1)}(t) = \displaystyle{\frac{{\sqrt{1 +
\frac{{n_{0}}^2\,R^2\,t^2}{a^2}}}}{t}}
\end{array},
\end{equation}
\begin{equation}\label{DTE0}\begin{array}{c}
D_{TE}^{(0)}(t) = \displaystyle{(2\,t +
\frac{2\,{n_{0}}^2\,R^2\,t^3}{a^2})^{-1}}
\end{array},
\end{equation}
\begin{equation}\label{DTE-1}\begin{array}{c}
D_{TE}^{(-1)}(t) = \displaystyle{\frac{{n_{0}}^2\,R^2\,t\,
     \left( 1 - \frac{{n_{0}}^2\,R^2\,t^2}{4\,a^2} \right) }{2\,
     a^2\,{\left( 1 + \frac{{n_{0}}^2\,R^2\,t^2}{a^2} \right) }^
      {\frac{5}{2}}} - \frac{\left( {n_{0}}^2 + \frac{2}{t^2}
       \right) \,Z^2}{2\,t\,{\sqrt{1 + {n_{0}}^2\,t^2}}}}
\end{array},
\end{equation}
\begin{equation}\label{DTE-2}\begin{array}{c}
D_{TE}^{(-2)}(t) =
\displaystyle{\frac{a^6\,{n_{0}}^2\,R^2\,t}{2\,{\left( a^2 +
{n_{0}}^2\,R^2\,t^2 \right) }^4}
    \left( + \frac{5\,{n_{0}}^2\,t^2}{2} -
      \frac{{n_{0}}^4\,R^4\,t^4}{4\,a^4}  -
       \frac{6\,{n_{0}}^2\,R^2\,Z^2}{a^2} -\right.}\\ \\
  \displaystyle{ \left. -
      \frac{a^2\,Z^2}{{n_{0}}^2\,R^2\,t^4} -
      \frac{4\,Z^2}{t^2}  - \frac{4\,{n_{0}}^4\,R^4\,t^2\,
         Z^2}{a^4} - \frac{{n_{0}}^6\,R^6\,t^4\,Z^2}{a^6} - 1  \right)}
\end{array}
\end{equation}
and
\begin{equation}\label{DTE-3}\begin{array}{c}
D_{TE}^{(-3)}(t) = \frac{{n_{0}}^2\,R^2\,t\,
     \left( 64\,a^6 - 560\,a^4\,{n_{0}}^2\,R^2\,t^2 +
       456\,a^2\,{n_{0}}^4\,R^4\,t^4 -
       25\,{n_{0}}^6\,R^6\,t^6 \right) }{128\,
     {\left(1 + \frac{{n_{0}}^2\,R^2\,t^2}{a^2}\right) }^{11/2}\,
    } + \\ \\
  \frac{Z^2\,\left( 16\,Z^2 + 56\,{n_{0}}^2\,t^2\,Z^2 +
       6\,{n_{0}}^8\,t^8\,Z^2 +
       3\,{n_{0}}^6\,t^6\,\left( t^2 + 12\,Z^2 \right)  -
       2\,{n_{0}}^4\,\left( t^6 - 35\,t^4\,Z^2 \right)
       \right) }{16\,t^5\,{\left( 1 + {n_{0}}^2\,t^2 \right)
         }^{\frac{7}{2}}} .
\end{array}
\end{equation}

\bigskip

Similarly, for the TM modes the functions $D_{TM}^{(k)}(t)$, $k =
1,...,-3$, appearing in the expansion $ D_{\nu}^{TM}(t)$, are
given by
\begin{equation}\label{DTM1}\begin{array}{c}
D_{TM}^{(1)}(t) = \displaystyle{\frac{{\sqrt{1 +
\frac{{n_{0}}^2\,R^2\,t^2}{a^2}}}}{t}},
\end{array}
\end{equation}
\begin{equation}\label{DTM0}\begin{array}{c}
D_{TM}^{(0)}(t) = \displaystyle{\frac{-1}{2\,\left( t +
\frac{2\,{n0}^2\,R^2\,t^3}{a^2} \right) }},
\end{array}
\end{equation}
\begin{equation}\label{DTM-1}\begin{array}{c}
D_{TM}^{(-1)}(t) = \displaystyle{\frac{-\left( {n_{0}}^2\,R^2\,t\,
       \left( 8\,a^2 + {n_{0}}^2\,R^2\,t^2 \right)  \right) }{8\,
     {\left( a^2 + {n_{0}}^2\,R^2\,t^2 \right) }^2\,
     {\sqrt{1 + \frac{{n_{0}}^2\,R^2\,t^2}{a^2}}}} -
  \frac{\left( 2 + {n_{0}}^2\,t^2 \right) \,Z^2}
   {2\,t^3\,{\sqrt{1 + {n_{0}}^2\,t^2}}}},
\end{array}
\end{equation}
\begin{equation}\label{DTM-2}\begin{array}{c}
D_{TM}^{(-2)}(t) = \displaystyle{\frac{{n_{0}}^2\,R^2\,t\,
     \left( 10 - \frac{10\,{n_{0}}^2\,R^2\,t^2}{a^2} +
       \frac{{n_{0}}^4\,R^4\,t^4}{a^4} \right) }{8\,a^2\,
     {\left( 1 + \frac{{n_{0}}^2\,R^2\,t^2}{a^2} \right) }^4} +
  \frac{Z^2}{2\,t^3}}
\end{array}
\end{equation}
and
\begin{equation}\label{DTM-3}\begin{array}{c}
D_{TM}^{(-3)}(t) = \frac{-\left( {n_{0}}^2\,R^2\,t\,
       \left( 176\,a^6 - 784\,a^4\,{n_{0}}^2\,R^2\,t^2 +
         480\,a^2\,{n_{0}}^4\,R^4\,t^4 -
         23\,{n_{0}}^6\,R^6\,t^6 \right)  \right) }{128\,a^8\,
     {\left( 1 + \frac{{n_{0}}^2\,R^2\,t^2}{a^2} \right) }^
      {\frac{11}{2}}}+ \\ \\
      + \frac{t^2\,
      \left( 2 + {n_{0}}^2\,t^2 \right) \,
      \left( 4 + 12\,{n_{0}}^2\,t^2 +
        3\,{n_{0}}^4\,t^4 \right) \,Z^2 +
     2\,{\left( 1 + {n_{0}}^2\,t^2 \right) }^2\,
      \left( 8 + 12\,{n_{0}}^2\,t^2 +
        3\,{n_{0}}^4\,t^4 \right) \,Z^4}{16\,t^5\,
     {\left( 1 + {n_{0}}^2\,t^2 \right) }^{\frac{7}{2}}}.
\end{array}
\end{equation}

\end{document}